\documentclass[twocolumn,showpacs,preprintnumbers,prl]{revtex4} 
\usepackage{graphics}
\usepackage{psfrag}
\begin{document}
\title{On adaptability and ``intermediate phase" in randomly connected networks }
\author{
J.~Barr{\'e}\thanks{jbarre@cnls.lanl.gov}, A.~R.~Bishop,
T.~Lookman, A.~Saxena}
\affiliation{Theoretical Division, Los Alamos National Laboratory, 
Los Alamos NM 87545, USA}

\date{\today}

\begin{abstract}
  We present a simple model that enables us to analytically
  characterize a floppy to rigid transition and an associated
  self-adaptive intermediate phase in a random bond network. In this
  intermediate phase, the network adapts itself to lower the stress
  due to constraints. Our simulations verify this picture. We use these
  insights to identify applications of these ideas in computational
  problems such as vertex cover and K-SAT.
\end{abstract}

\maketitle

The concepts of rigidity and a rigidity
transition~\cite{Phillips79,Thorpe83} have been successfully applied
to the study of network glasses.  The bonds in a network are
considered as constraints such that at low connectivity there are more
degrees of freedom than constraints, so that the network is flexible,
or floppy. At high connectivity, there are more constraints than
degrees of freedom, and the network is said to be rigid and stressed.
The rigidity transition lies in between and Thorpe \emph{et
  al.}~\cite{Thorpe00} have recently suggested the idea of
\emph{adaptability} of a network to avoid stress, which has led to the
discovery of an \emph{intermediate phase}, between the usual floppy
and rigid ones. The existence of this phase has been confirmed by
numerical studies \cite{Thorpe00}, analysis of finite size
clusters~\cite{Micoulaut} and by experiments~\cite{Boolchand}.  What
has been lacking is an appropriate model of the intermediate phase
that allows for analytical calculations and provides insight.
Furthermore, the ingredients at the origin of the intermediate phase
are quite generic: adaptability of the underlying network undergoing a
transition, so as to avoid stress. This suggests that such
intermediate phases may be relevant in many different fields. For
instance, links between rigidity theory and computational phase
transitions have already been suggested~\cite{Monasson}, but not
examined further than an analogy. The insight gained from the minimal
model we discuss here will demonstrate how an intermediate phase can
arise in computational problems such as the vertex cover
problem~\cite{Weigt} or K-SAT~\cite{Mezard}.

We will first present and solve a simple random bond model 
for the rigidity transition. We then introduce the
possibility of adaptation of the network. Through mean-field
calculations, we demonstrate the presence of an intermediate phase and
study its properties. The calculations are quantitatively confirmed by
Monte Carlo simulations. We then outline how the results could
be extended to computational phase transitions.

\begin{figure}
%\psfrag{x3}{{\huge $x_3$}}
%\psfrag{flop}{\hskip -4cm {\huge floppy modes and cluster sizes}}
\resizebox{0.45\textwidth}{.22\textheight}{\includegraphics{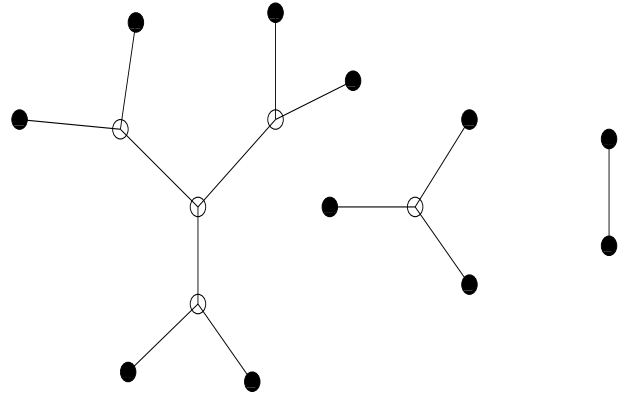}}
%\vskip -1truecm
\caption{Example of a network with $N_1=11$, $N_3=5$, so that
  $x_3=0.3125$. It has $N_{11}=1$,
$N_{13}=9$, $N_{33}=3$, $N_3^{(0)}=1$, $N_3^{(1)}=3$,
$N_3^{(2)}=0$, $N_3^{(3)}=1$ (see text). The two subgraphs on the
right are rigid; the one on the left has three internal degrees of
freedom: the rotations around the $3-3$ bonds. There is no stress in
the network.}
\label{fig:network}
\end{figure}

\begin{figure}
\psfrag{x3}{{\huge $x_3$}}
\psfrag{flop}{\hskip -4cm {\huge floppy modes and
      cluster sizes}}
\resizebox{0.48\textwidth}{.25\textheight}{\includegraphics{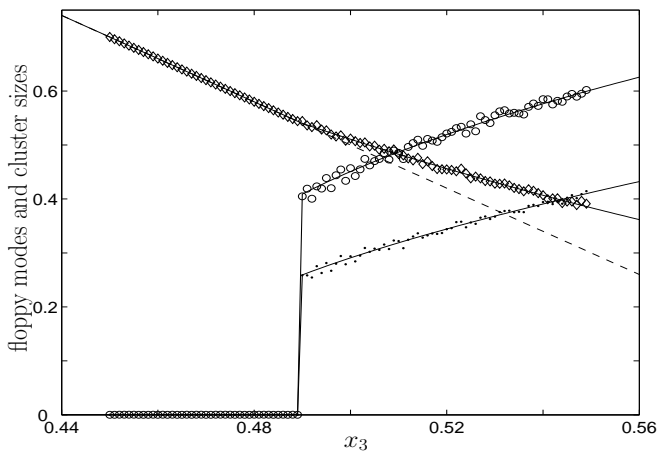}}
%\vskip -1truecm
\caption{Comparison of analytical calculations (solid lines) with
  numerics (symbols). The diamonds represent the number of floppy modes per
  atom; the circles (dots) the fraction of the network in
  the percolating rigid (stressed) cluster. The dashed line is
  the constraint counting estimate for the number of floppy modes. The
  numerics are performed on networks of $4\times10^4$ atoms.}
\label{fig:rig_trans}
\end{figure}

Very few non-trivial exactly solvable models are available for the
standard rigidity transition. To our knowledge, the randomly bonded
models, equivalent to Bethe-like lattices~\cite{Moukarzel}, are the
only ones available. The first idea of this paper is therefore to
introduce \emph{adaptability} in the randomly bonded models to obtain
a description of the complete phase diagram, including the
intermediate phase. We consider a simple model for the rigidity
transition. Our network consists of N atoms, $N_3=Nx_3$ of which are
3-fold coordinated, and $N_1=Nx_1$ are 1-fold coordinated; we consider
bond stretching as well as bond bending constraints. Atoms are bonded
randomly, disregarding space: this leads locally to a tree-like
network, which allows for analytical calculations~\cite{Moukarzel}; an
example with $N=16$ atoms is given Fig.~\ref{fig:network}. Our aim is
not to accurately describe network glasses, rather to seek a
simplified description of the salient features.  We invoke the
following constraint counting argument: each atom has a priori $3$
degrees of freedom (the network is embedded in 3-dimensional space),
each 1-atom brings $1/2$ constraint ($1$ bond stretching shared with
its neighbor and no bond bending), and each 3-atom brings $9/2$
constraints ($3/2$ bond stretching and $3$ bond bending). If no
constraint is redundant, the number of unconstrained degrees of
freedom (or floppy modes) is given by the Maxwell
estimate~\cite{Phillips79,Thorpe83}
\begin{equation}
N_{flop}=3N-\frac{Nx_1}{2}-\frac{9Nx_3}{2}~.
\label{eq:maxcounting}
\end{equation}
This yields the estimate for the transition point $x_3^{\ast}\simeq
0.625$, for which the mean number of constraints per atom equals~$3$,
the number of degrees of freedom \cite{footnote}.  The calculation can
be carried out exactly using a method developed in~\cite{Chubinsky} or
by adapting the cavity formalism originally devised for spin
glasses~\cite{Cavity}. The cavity method has been used
recently~\cite{Mezard03} to study a computational phase transition:
its use is thus the first formal link, beyond a simple analogy,
between rigidity theory and the field of computational phase
transitions. These calculations confirm the presence of a rigidity
transition, first order in this case, at a concentration of 3-atoms
(i.e. atoms with three bonds) $x_3^{\ast}=0.490$; see
Fig.~\ref{fig:rig_trans}.
%Let us note that the
%presence of 1-fold coordinated atoms randomly bonded implies the
%presence of isolated dimers, and more generally of isolated
%subnetworks. 
At the rigidity transition, a macroscopic rigid
cluster appears. In the rigid phase, some constraints cannot be
fulfilled, and stress is present.

As stress costs energy, it is natural to assume that the network
will try to adapt itself to avoid it. To make this idea more precise,
let us define a stress energy for a given configuration of the network,
equal to the number of redundant constraints, that is constraints
that cannot be fulfilled. This energy is of course zero in the floppy
phase where all constraints can be accommodated, and non-zero in
the rigid phase, as shown on Fig.~\ref{fig:rig_trans}. Constraint
counting gives
\begin{equation}
N_{flop}=3N-\frac{1}{2}N_1-\frac{9}{2}N_3+N_{red}~,
\label{eq:constaintcounting}
\end{equation}
where $N_{flop}$ is the number of floppy modes in the network,
$N_{red}$ is the number of redundant constraints, $3N$ is the a priori
number of degrees of freedom, and $(N_1+9N_3)/2$ is the total number
of constraints. Thus, for a fixed concentration of 3-atoms, counting
$N_{flop}$ amounts to the same as counting $N_{red}$. Consequently, in
the following, we will use indifferently
\begin{equation}
H=N_{flop}~\mbox{or}~H=N_{red}~.
\label{eq:hamiltonian}
\end{equation}
%A comment is in
%order here. It would be logical to assign an entropy gain to floppy
%modes, thus an entropy gain to a redundant constraint. For certain
%temperatures however, the energy loss would dominate. In the
%following, we assume that a redundant bond is unfavored.
We assume in addition that the network is adaptive: i.e. while $N_1$
and $N_3$ are fixed, some bonds can be rewired to decrease the
energy~(\ref{eq:hamiltonian}), thus moving the network away from the
randomly bonded case. The standard rigidity transition described above
thus corresponds to the infinite temperature case. At finite
temperature, new phenomena arise (below).

To realize the calculations, it is crucial to devise a means to
measure the degree of organization or randomness of the network. At
the crudest level, this is provided by the number of bonds between two
$1$-atoms, one $1$-atom and one $3$-atom, and two $3$-atoms,
respectively termed $N_{11},N_{13},N_{33}$, see
Fig.~\ref{fig:network}. In the randomly bonded case, a simple
calculation shows
\begin{eqnarray}
N_{11} &=& \frac{N}{2} \frac{x_1^2}{x_1+3x_3} =N_{11}^{\ast}~, \\
N_{13} &=& \frac{N}{2} \frac{6x_1x_3}{x_1+3x_3} =N_{13}^{\ast}~, \\
N_{33} &=& \frac{N}{2} \frac{9x_3^2}{x_1+3x_3} =N_{33}^{\ast}~. 
\label{eq:N11N13N33}
\end{eqnarray} 
At finite temperature, $N_{ij}$ can now deviate from its random value
$N_{ij}^{\ast}$, with the constraints $2N_{11}+N_{13}=N_1$ and
$2N_{33}+N_{13}=3N_3$. The level of organization of the network, at
this crude one bond level, is thus determined by a single parameter,
which we take to be $a=N_{11}/N_{11}^{\ast}$. Fixing $a$ fixes all
$N_{ij}$; $a=1$ corresponds to the random bonding case, and $a\neq
1$ denotes an adaptation of the network. Assuming that there are no
correlations in the network beyond the one bond level, the
Chubinsky~\cite{Chubinsky} or the cavity methods allow exact calculations
of the energy $H$ as a function of~$x_3$ and~$a$. It turns out
that the lower $a$, the higher is the rigidity threshold for $x_3$.
This is intuitive: decreasing the number of dimers of two 1-atoms at
fixed $x_3$ decreases the mean connectivity of the main network, which
is then less likely to be rigid. At fixed $a$, knowing the $N_{ij}$,
it is a simple combinatorial problem to compute the configurational
entropy of the network; $S(x_3,a)=Ns(x_3,a)=\ln \Omega$, with
\begin{equation}
\Omega(N_{11},N_{13},N_{33})=
\frac{N_1!~(3N_3)!}{2^{N_{11}}2^{N_{33}}~N_{11}!~N_{13}!~N_{33}!}. 
\label{eq:entropy}
\end{equation}

\begin{figure}
\psfrag{a}{{\huge a}}
\psfrag{elibre}{\hskip -4cm{\huge energy, entropy, free energy}}
\resizebox{0.48\textwidth}{!}{\includegraphics{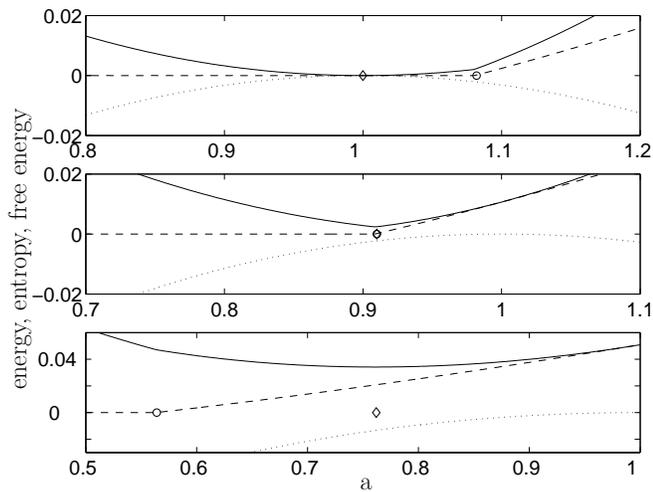}}
%\vskip -1truecm
\caption{Energy (dashed line), entropy (dotted line) and free energy
  (solid line) as a function of the parameter $a$ for different $x_3$; 
  from top to bottom, $x_3=0.48$ (floppy phase), $x_3=0.5$ (intermediate
  phase), $x_3=0.53$ (rigid phase). The circle indicates the rigidity
  transition varying $a$ at fixed $x_3$; the diamond indicates the free
  energy minimum; both coincide in the intermediate phase.}
\label{fig:thermo}
\end{figure}

\begin{figure}
%\psfrag{0.5}{\hskip -0.5cm \huge $0.5$}
%\psfrag{1.0}{\huge $1$}
%\psfrag{0.44}{\huge $0.44$}
%\psfrag{0.48}{\huge $0.48$}
%\psfrag{0.52}{\huge $0.52$}
%\psfrag{0.56}{\huge $0.56$}
%\psfrag{er}{\huge $0.6$}
\psfrag{x3}{\Huge $x_3$}
\psfrag{nflop}{ \hskip -6cm{\huge Number of floppy modes and parameter $a$}}

\resizebox{0.48\textwidth}{.25\textheight}{\includegraphics{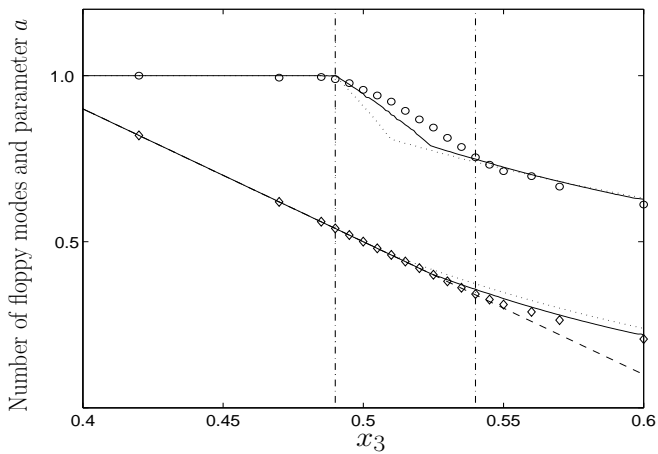}}
%\vskip -1truecm
\caption{Comparison between MC simulations (circles for the $a$
  parameter, diamonds for the number of floppy modes per atom) and
  analytical results (dotted lines correspond to the one bond
  correlation level and solid lines to the two bond correlation
  level). The two dot-dashed vertical lines indicate the location of
  the two phase transitions. The dashed line indicates the Maxwell
  estimate for the number of floppy modes, see
  eq.~(\ref{eq:maxcounting}). MC simulations involve 5000 or
  20000 atoms, for $10^6$ MC steps.}
\label{fig:anandnum}
\end{figure}

The conclusion of the calculation now only amounts to minimizing with
respect to the parameter~$a$ the free energy $H(x_3,a)-TS(x_3,a)$, at
given $T$ and $x_3$. This step is sketched in Fig.~\ref{fig:thermo}.
What occurs is again intuitively clear: for $x_3<x_3^{\ast}$, the
random network is unstressed, thus entropic and energetic
contributions are optimized for $a=1$, and the energy is zero.  For
$x_3>x_3^{\ast}$, random bonding has a non-zero energy; however, for
small enough $x_3$, decreasing $a$ to avoid stress energy costs little
entropy, and is favorable. This is the intermediate phase. Throughout
this phase, the system adapts itself to stay exactly on the verge of
stress, so that the entropic cost is minimal and there is no stress.
For $x_3$ large enough, decreasing $a$ further to avoid stress costs
too much entropy, and the appearance of some stress is favored.  This
happens at a second transition~$x_3=x_3^{\ast\ast}$. The results are
summarized in Fig.~\ref{fig:anandnum}. To test this proposed scenario,
we performed Monte Carlo simulations of the system as follows: at each
step, we pick a pair of bonds and rewire them, accepting the move
according to the Metropolis algorithm with
energy~(\ref{eq:hamiltonian}), and temperature chosen to be $T=5$.
Choosing another temperature does not induce any qualitative change;
only the width of the intermediate phase is affected: the lower the
temperature, the wider the intermediate phase. In the limit of a
non-adaptive network $T\to\infty$, the intermediate phase disappears.
The network is analyzed, when needed, using the pebble game
algorithm~\cite{pebble}, which gives access to the number of floppy
modes, and thus the energy~(\ref{eq:hamiltonian}), and provides the
decomposition of the network into rigid and stressed clusters. We
performed simulations on $N=5000$ and $N=20000$ systems, for $10^6$
Monte Carlo steps. Longer test runs did not show any significant
differences.  Results for the parameter $a$ and the number of floppy
modes, Fig.~\ref{fig:anandnum}, show a qualitative agreement with the
simple calculation above: a floppy phase with $a=1$ at low $x_3$, a
stressed phase at high $x_3$ and an intermediate unstressed but
self-adapted $a\neq 1$ phase.

The agreement is quantitatively not very good: the assumption of
retaining only one bond correlations most likely breaks down. The
above calculation can indeed be refined to take into account longer
range correlations. The first steps in this direction consist in
counting not only the bonds as above, but also the paths along two
bonds which are of four types: from $1$ to $3$ to $1$, from $1$ to $3$
to $3$, from $3$ to $3$ to $1$, and from $3$ to $3$ to $3$. This is
equivalent to counting $N_3^{(0)},N_3^{(1)},N_3^{(2)}, N_3^{(3)}$, the
number of 3-atoms that are linked respectively with zero, $1$, $2$ and
$3$ other 3-atoms.  Since the two equations
\begin{eqnarray}
3N_3^{(0)}+2N_3^{(1)}+N_3^{(2)}+2N_{11} &=& N_1 \\
N_3^{(0)}+N_3^{(1)}+N_3^{(2)}+N_3^{(3)} &=& N_3
\label{eq:relations}
\end{eqnarray}
must be satisfied, describing the network at the level of two bond
correlations requires the introduction of two new parameters, defined
as follows:
\begin{eqnarray}
\alpha_0 &=& \frac{N_3^{(0)}}{N_3^{(0)\ast}} \\
\alpha_1 &=&\frac{N_3^{(1)}}{N_3^{(1)\ast}}~, 
\label{eq:defalpha}
\end{eqnarray}
where $N_3^{(i)\ast}$ is the number $N_3^{(i)}$ expected without two
bond correlations. Thus, any $\alpha_i\neq 1$ shows the presence of
two bond correlations. Using the cavity method, one can calculate the
energy $H(x_3,a,\alpha_0,\alpha_1)$. As above with the parameter $a$,
it turns out that the lower $\alpha_0$ and $\alpha_1$ are, the less
stress-prone is the network. Evaluating the entropy
$S(x_3,a,\alpha_0,\alpha_1)$ is still a simple combinatorial problem.
The results are plotted in Fig.~\ref{fig:anandnum}, and are much
closer to the numerical results. Due to the simplicity of the model,
we were able to make the calculations up to three bonds correlations;
they confirm the trend toward the numerical results, but do not
completely coincide with them in the intermediate phase. This leads to
the important conclusion that medium or long range correlations are
important in the network, especially in the intermediate
phase~\cite{Micoulaut}.

\begin{figure}
\psfrag{x3}{\huge $x_3$}
\psfrag{proba}{\hskip -4cm{\huge probability of a percolating cluster}}
\resizebox{0.48\textwidth}{.25\textheight}{\includegraphics{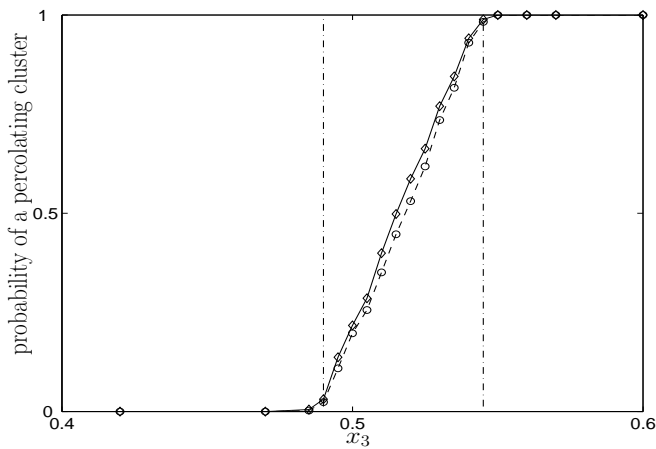}}
%\vskip -1truecm
\caption{Diamonds (circles): probability that a rigid
  (stressed) cluster percolates the entire sample, as a function
  of $x_3$. The runs were made with $N=5000$ or $N=20000$ atoms, for
  $10^6$ MC steps. Solid and dashed lines are guides for the eyes;
  dot-dashed lines indicate the location of the two phase
  transitions.}
\label{fig:proba}
\end{figure}

Monte Carlo simulations also provide the opportunity to examine the
different phases in greater detail. The most important question
concerns the presence or absence of percolating rigid or stressed
clusters. In the floppy phase, there is no percolating rigid or
stressed cluster, whereas in the stressed phase rigidity and stress
always percolate. The interesting case is the intermediate phase: our
simulations show that the probability to find a rigid or stressed
percolating cluster varies smoothly from $0$ to $1$, see
Fig.~\ref{fig:proba}. The probability to find a stressed cluster is
always slightly lower than the probability to find a rigid one. Thus,
while in the intermediate phase the system lies right on the boundary
between rigid and floppy, the probability to be on one side or another
of the boundary goes from $0$ to $1$. Although strictly speaking the
model gives access only to Monte Carlo sampling of the phase space, we
can attempt to infer a dynamical picture of the intermediate phase,
namely a percolating rigid cluster disconnecting and reconnecting
elsewhere through local connections.

From the detailed analysis above, the ingredients leading to the
intermediate phase appear to be clear: an underlying first order phase
transition, and the possibility of adaptation of the network, giving
rise to an entropy competing with the energy associated with the
underlying transition.  One thus expects the same phenomena for a
variety of computational phase transitions. Consider the vertex cover
problem as an example.  The problem is as follows: given a network of
size $N$, is it possible to find a subset of vertices of size $xN$,
such that all edges are adjacent to this subset? Such a subset is
called a vertex cover. If the network structure is random and fixed,
this is always possible at low connectivity, and becomes impossible at
high enough connectivity~\cite{Weigt}, through a phase transition. If
one defines the energy of a configuration as the number of edges not
adjacent to the subset chosen, and allows for a reorganization of the
network, an intermediate phase with a finite probability of finding a
vertex cover is likely to appear. This should apply also to
satisfiability problems of the K-SAT type~\cite{Mezard}.
%(maybe not 2-SAT, since the underlying
%transition is then second order~\cite{Monasson}). 
More generally, as
most networks found in physics, social science or biology, are
adaptive, the intermediate phase described in this work is
likely to be encountered in a wide variety of situations. Future work
will elaborate the thermodynamic and dynamical signatures of the
intermediate phase, including expectations of filamentary geometry and
glassiness.\\

We would like to thank Don Jacobs, Mykyta Chubinsky and Michael
Thorpe for permission to use the software Pebble-3D. Motivational
discussions with Jim Phillips are gratefully acknowledged. Work at Los
Alamos National Laboratory is supported by the US Department of Energy.


\begin{thebibliography}{99}
\bibitem{footnote} This argument can actually be refined as shown
  in~\cite{Chubinsky}; it then gives a much better result.
\end{thebibliography}
\end{document}